\documentclass{evn2002}
\usepackage{graphicx}
\begin{document}
   \title{Centimeter-band Variability in GPS Sources}

   \author{M. F. Aller, H. D. Aller, P. A. Hughes, \& R. M. Plotkin}

   \institute{Department of Astronomy, University of Michigan, 
   Ann Arbor, MI 48109 USA}

   \abstract{Monitoring results are presented for the
Stangellini 1 Jy GHz-peaked-spectrum source sample, illustrating that several
members exhibit variability in total flux and/or linear polarization over
timescales of order a decade. The variability occurs while the spectrum, based
on the integrated fluxes, remains steep and characteristic of a transparent source.
Total flux variability is unexpected  in view of recent VLBI observations indicating 
no or hidden cores in several members. However, both the variability, and the 
detection of circular polarization in one class member, argue for the presence of
opacity in some portion of the radio jet. 
   }

   \maketitle
%
%________________________________________________________________

\section{Introduction}

To investigate the total flux density and polarization properties of
GHz-peaked sources, we commenced systematic observations of the
1 Jy Stangellini (\cite{stan98a}) sample with the University of Michigan
Radio Astronomy Observatory (UMRAO) 26-meter telescope in November 1999. 
The class of objects is of particular interest because they are believed
to be young sources, primarily based on their small sizes, or, alternatively,
sources confined by dense gas in their ambient medium (\cite{odea}).
Expected class properties from prior VLA and VLBI observations (e.g. \cite{odea})
are low level variability and low ($<1\%$) fractional linear polarization.
Recent VLBA observations at 15 GHz suggest no or hidden cores (\cite{stan01}).

The selection criteria for the sample are:
\newline 1) declination $\delta > -25^{\circ}$, galactic latitude $|b|>10^{\circ}$;
\newline 2) S$> 1$ Jy at 5 GHz:
\newline 3) Turnover frequency in the range $0.4\leq\nu_{to}\leq 6$ GHz;
and spectral index $\alpha>0.5$ in the high frequency (presumably optically
thin) part of the spectrum ($S_{\nu}\propto\nu^{-\alpha})$.
\newline By optical class the objects are a mix of quasars and galaxies.
By radio morphology 1/3 of the sample are known or suspected compact symmetric
objects (cso), a group identified as exhibiting low variability (\cite{fass}).

Thirty-two of the 33 sample members are within the declination limits of the
Michigan 26-meter paraboloid. These objects have been observed at least
tri-monthly at 4.8, 8.0, and 14.5 GHz for total flux density and linear
polarization since November 1999. Eighteen of the sources have been observed
for a decade or more with sufficient sampling to identify variability, 
if present; limited data with ad hoc sampling are available for 9 additional
objects; and only 6 sources had not been observed by us prior to November 
1999. Twelve objects are currently members of the 2cm VLBA survey MOJAVE
(\cite{kell}), providing important complementary information on pattern 
speeds and structural changes. Preliminary results for some sample members
indicate relatively slow component speeds (Lister private communication;
see also http://www.cv.nrao.edu/2cmsurvey/).

In  Figure 1 we compare fractional linear polarization for the GPS sample
and for the Pearson-Readhead (hereafter PR) sample (\cite{pearson}), a flux
and position limited sample; the latter is taken for comparison because it
is believed to be representative of extragalactic objects, and insensitive to 
most selection effects. We have systematically been observing it since 
1984 (\cite{aller}). Only galaxies and QSOs are included
for the PR sample since there are no BL Lacs with GPS spectra. Comparison of
the distributions confirms that that on average the 4.8 GHz polarization 
is low in GPS sources relative to the general population of
extragalactic objects; a KS test gives a low probability of 0.000034 that
these distributions are drawn from the same parent population.

\begin{figure}
\includegraphics[width=10cm]{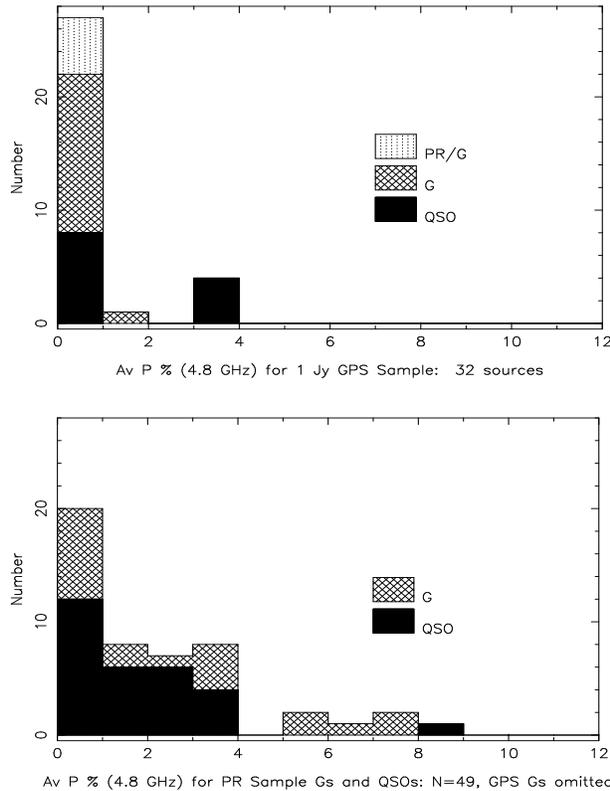}
\caption{Average polarization from time-averages of the daily measurements of
the Q and U Stokes parameters. Top panel: GPS sources; bottom panel: Pearson-Readhead
sources. P has been corrected for bias due to random noise (\cite{ward74}). The 
five sources common to both samples are omitted from the distribution in the
bottom panel.
   } 
\label{fig1}
    \end{figure}

In contrast to the abnormally low fractional polarizations at 4.8 GHz
however, several sources show fractional polarization at 14.5 GHz at 
levels comparable to those in AGNs; in several these are clearly time-variable
with time scales of order several years to a decade. As an example, we show in Figure 2, 
the total flux density and linear polarization light curves for the galaxy, 1358+624, 
a source classed as a cso by radio morphology. While there is little to no flux
variability as expected for objects of this type, a long-term outburst in linear
polarization occurred at both 8 and 14.5 GHz during the mid to late 1980s. We
find similar behavior, low amplitude variability in total flux and variability
in linear polarization at 14.5 and tentatively at 8.0 GHz in the suspected cso, 1117+146.

Plausible mechanisms discussed in the literature to explain low fractional
linear polarizations are (\cite{stan98b}) Faraday depolarization in the
radio-emitting region, or a tangled magnetic field along the line of sight.
A third possibility, beam  depolarization, has recently been rejected 
(\cite{stan01}). We believe that a tangled magnetic field would not 
readily account for the frequency-dependent differences in linear 
polarization we find in our data. 

\begin{figure}
\centering
\includegraphics[angle=-90,width=9cm]{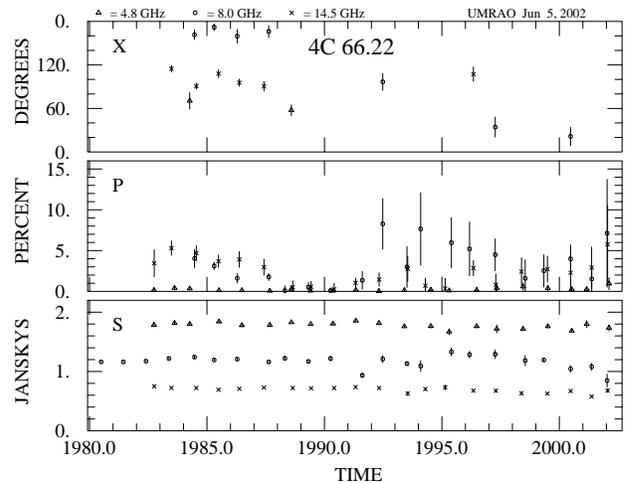}
\caption{Light curves for 4C 66.22 (1358+624) based on yearly averages of
the UMRAO data. From bottom to top: total flux density, fractional linear
polarization, and position angle of the electric vector of the polarized
emission (EVPA). The data at 4.8 GHz are denoted by triangles, at 8 GHz by
circles,and at 14.5 GHz by crosses. This convention is adopted in subsequent
figures.}
\label{fig2}
 \end{figure}

In Figure 3 we show light curves for the QSO 1127-145, a source which has
exhibited variability in both total flux and linear polarization. A goal
of our program has been to determine whether the GPS spectrum is maintained
over decades, or whether for some sources this spectral character is a short-term
phase. In the case of 1127-145, the source maintained its GPS spectral shape for
nearly 20 years, but in 1999 a large outburst commenced at 14.5 GHz, and the spectrum
flattened.  Thus, based on its current spectrum, it would not have been classed 
as a GPS source.

\begin{figure}
\centering
\includegraphics[angle=-90,width=9cm]{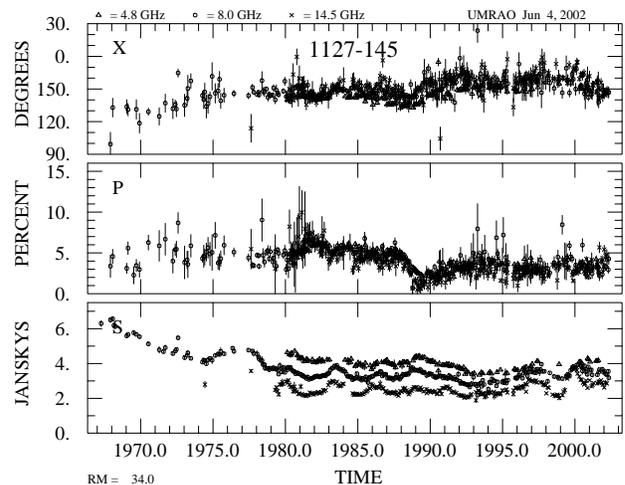}
    \caption{Monthly averages of the data for 1127-145. The polarization
has been corrected for Faraday rotation assuming the low value of 34 rad/m$^2$
given in \cite{rusk}. The EVPAs in the top panel do not show a $\lambda^2$ 
separation, the signature of Faraday rotation.}
\label{fig3}
    \end{figure}

To investigate the nature of the variability in this source, we show
in Figure 4 a structure function based on the data shown in Figure 3. 
Structure functions are useful tools providing two pieces of information.
The slope of the linear portion gives information on the noise process 
responsible for the variability. It is typically found to be $\sim$1 in
AGNs (\cite{hughes}) a result indicative of shot noise. The turnover
gives a measure of the characteristic time scale and is typically of
order 2 years, with some spread from source to source. In 
the structure function for this source we find a slope
of $0.57\pm0.18$, which does not match {\it any} well-studied, ideal
noise processes, and a relatively long, frequency-dependent characteristic
variability time scale ranging from 4.1 years at 4.8 GHz to over a decade
at 8 and 14.5 GHz. The behavior of this structure function illustrates that
the nature of the variability is both unusual and complex in this source.

\begin{figure}
\centering
\includegraphics[angle=-90,width=9cm]{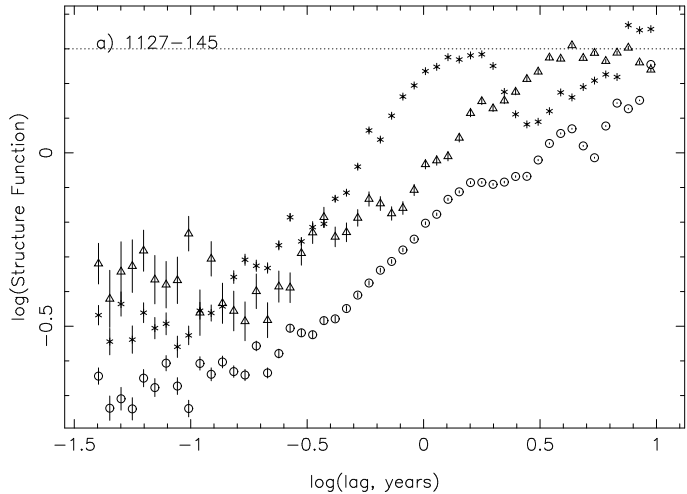}
\caption{First-order structure function for 1127-145 based on monthly
 averages of the data. The dotted line near the top marks $2\sigma^2_{signal}$.}
\label{fig4}
\end{figure}

In Figure 5 we show light curves for a second variable source, 2134+004,
which has been well-studied with VLBI (e.g. \cite{paul}). Rotation measure mapping
(\cite{taylor}) has identified significant Faraday rotation in
this source which is highly position-dependent. Our EVPAs have been 
corrected for Faraday rotation adopting Taylor's value for component A, 
one of two nearly-identical
unresolved components. This correction brings the EVPAs at the 3
frequencies into agreement at Taylor's VLBA epoch (1998.58), but at other
epochs a wavelength-dependent separation remains which most likely is
due to a combination of time-variable opacity and time-variable Faraday
rotation. We are, unfortunately, unable to unravel these effects
with our single dish measurements.

  \begin{figure}
   \centering
 \includegraphics[angle=-90,width=9cm]{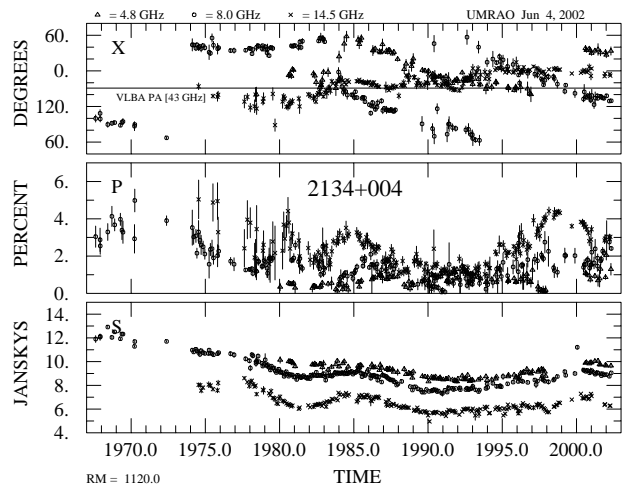}
 \caption{Monthly averages of the data for 2134+004. The polarization had been
corrected for Faraday rotation assuming a time-independent rotation measure of
1120 rad/m$^2$. An ambiguity of 180$^\circ$ in our EVPA determinations produces
the apparent discontinuities in the EVPA light curve. The line in the top panel
marks the orientation of the VLBI structural axis determined from data at 43 GHz
(\cite{list}).}    
\label{fig5}
    \end{figure}

In Figure 6 we show the light curves for OI~363, a QSO for which our sampling is
irregular. While the observations cannot be used to identify the details of the
variability, they clearly reveal that the source is variable in both total flux density and
fractional linear polarization, and the data are consistent with a variability time scale
of order a decade. Note that the GPS spectral shape in maintained during the observing
window of two decades.

\begin{figure}
\centering
\includegraphics[angle=-90,width=9cm]{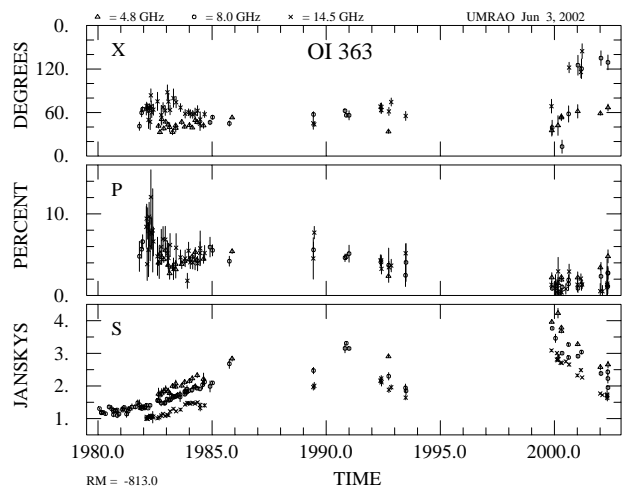}
\caption{Daily averages of the data for OI 363 (0738+313). The polarization has been
corrected for Faraday rotation assuming the rotation measure given in \cite{stan98a}
Note that the fractional linear polarization exceeded 10\% at 14.5 GHz in the
early 1990s but is now only a few percent.}
\label{fig6}
\end{figure}

In Figure 7 we show data for a poorly sampled source which has only been observed with
regularity since the start of the program. However, even with this limited
sampling, the ordered changes in total flux density since late 1999 are consistent
with the presence of low level variability. While the linear polarization is only 
of order a few percent, this is a source in which we have recently detected 
relatively high levels of circular polarization ($-0.71\pm0.07\%$)
in our single dish measurements at 4.8~GHz; CP with a comparable
magnitude and sign has also been detected in this source by \cite{homan} ($-0.46\pm0.06\%$) 
using VLBA observations obtained in December 1996, a time period in which
we, unfortunately, have no data. The most plausible mechanism for the production
of the circular polarization is linear-to-circular conversion in a partially-opaque
emitting region.

\begin{figure}
\centering
\includegraphics[angle=-90,width=9cm]{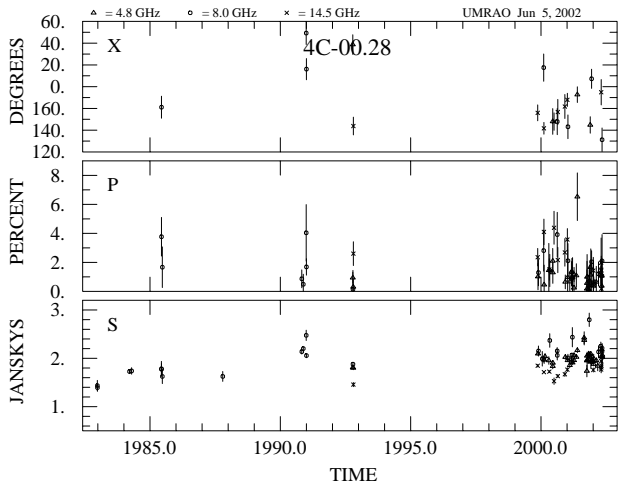}
\caption{Daily averages for the core-jet source, quasar
\hfil\break 4C-00.28(0743-006)}
\label{fig7} 
\end{figure}

%
   
%______________________________________________________________

\section{Conclusions}

Our main conclusions are:
\newline 1) Several sample members exhibit variability in total flux with
amplitudes comparable to those in AGNs. Of the 18 good-to-well observed sources,
10 show total flux density variations at some level. The characteristic time
scales of these variations are typically long compared to other UMRAO program sources.
\newline 2) In general the characteristic GPS spectrum is maintained over decades, 
including times when flux variability is present.
\newline 3) The  time-averaged fractional linear polarizations at 4.8 GHz are low
compared to results based on the Pearson-Readhead sample at 4.8 GHz, confirming
that this is a GPS class property. 
\newline 4) At 14.5 GHz, several sources exhibit higher fractional polarization than
at 4.8 GHz which in some sources is time-variable, reaching values as high as 
10$\%$;  2 confirmed or suspected cso sources exhibit extremely low level 
variability in total flux while exhibiting well-defined variations in fractional
polarization at 14.5~GHz, as illustrated in Figure 2.
\newline 5) The frequency-dependence in the EVPA in some sources is consistent with 
significant  Faraday rotation, which may be time-variable.

While the spectral shape is characteristic of a transparent source,
both the variability we find and the detection of circular polarization in one
class member are most easily explained by the presence of significant opacity 
in some portion of the jet flow. This hypothesis can best be tested by spectral
mapping of selected group members. Hopefully, longer-term observations will answer
the questions of whether variability is a property of most, if not all, sample
members which are not csos, and, more importantly, of whether the sample is, in fact,
a collection of physically-different objects joined together on the basis of their
(in some cases) evolving radio spectra.

\begin{acknowledgements}
UMRAO is operated with funds from the Department of Astronomy of the University of
Michigan.

\end{acknowledgements}

\end{document}